The Arrhenius Law Prefactor in Permalloy Mesoscale Systems

J. T. Delles[a,b], E. Dan Dahlberg[a,c]

[a]School of Physics and Astronomy, University of Minnesota, Minneapolis, Minnesota 55455

[b]corresponding author: delle011@umn.edu, [c]dand@umn.edu

**Abstract**

The Arrhenius equation was used to describe the dynamics of two-state switching in mesoscale, ferromagnetic particles. Using square, permalloy dots as an idealized two-state switching system, measurements of the prefactor of the Arrhenius law changed by 26 decades over barrier heights from 30 meV to 700 meV. Measurements of the prefactor ratios for a two well system revealed significant deviations from the common interpretation of the Arrhenius law. The anomalous Arrhenius prefactors and the prefactor ratios can be fitted to a modified model that includes entropic contributions to two-state transitions. Similar considerations are likely for the application of the Arrhenius law to other mesoscale systems.

## INTRODUCTION

Since Arrhenius modeled the escape from a single well in 1889 [1, 2] and later extended by Kramers in 1940 to two wells [3], there has been an extensive body of research involving transition rates that relied on the expression known as the Arrhenius law given by

$$\tau = \tau_o e^{U/k_bT} = f_o^{-1} e^{U/k_bT}, \qquad (1)$$

where $\tau$ is the average dwell time of a state, $\tau_o$ is a prefactor generally taken as the characteristic dwell time and $f_o$ is the inverse usually taken as an attempt frequency, $U$ is the energy barrier for the state, $k_b$ is the Boltzmann constant, and $T$ is the temperature. Kramers' extension included both overdamped and underdamped cases. Kramers was unable to unify the results from both regimes and became known as the Kramers' turnover problem. Although extensively used to model physical, chemical, and biological systems [4-6], an in-depth physical understanding is lacking at the mesoscale; the exponential part, referred to as the Boltzmann factor, is well established, but the prefactor is less well understood at the mesoscale. There have been numerous prefactor, $\tau_o$, theories in classical systems with various levels of damping [3,7], with multiple degrees of freedom [8-10], with various forms of noise [11], for quantum systems [12, 13], and for ferromagnetic systems[14]. Despite this, there has been limited quantitative research of $\tau_o$ [15-18] and it has been insufficient for modelling.

Brown extended Kramers' work to a superparamagnetic, nanoscale system using the Landau-Lifshitz-Gilbert equation as the basis for his work [13, 19]. He found the Arrhenius law prefactor for a ferromagnetic particle depends upon the magnetic energies of the particle, the magnetic damping, and the temperature of the particle. In addition to Brown, others have applied this work to systems of many degrees of freedom where the resulting prefactor depends on the ratio of the products of the natural frequencies associated with the curvature of the energy landscape at the bottom of the well and at the maximum, or saddle point [14, 19].

Some experiments that have been performed to explore two-state switching include individual ferromagnetic nanoparticles of various sizes [15], collections of noninteracting ferromagnetic nanoparticles [16], quantum tunneling in a Josephson junctions [17], and chemical systems [18] to name a few. None of these studies had sufficient experimental data to explore the physics of $\tau_o$ despite Kramers' two-state switching theory being extensively used for almost 80 years. The need for such a study is clear from the limited experimental evidence that the prefactor is more complex than expected [15, 20]

Here, we report an extensive study of 140 prefactors of the Arrhenius law derived from over 14,000 measured dwell times in mesoscale magnetic particles. Our measured prefactors range from $10^{-2}$s to the unphysically small values of $10^{-28}$s if $\tau_o$ is considered to be a characteristic dwell time, or the inverse, $f_o$, the characteristic attempt frequency. The independently determined barrier heights separating the two states range from 30 meV to 700 meV. We measured the individual dwell times for each well in the particle separately and determined each well's individual barrier height and prefactor. The previously mentioned models [3, 7-9, 11 12, 14] and others we have explored [20, 21] do not explain our observations. We did find that with a slight modification, the theory by Talkner [10] for $\tau_o$ that considers a multidimensional system in the limit of $U/k_bT \gg 1$, can replicate our results though there remain questions as to it being the full explanation. In what follows, we will discuss the particles, how they are made, and how the two-state switching is measured. We then discussed the theoretical modeling followed by an analysis of the data.

**EXPERIMENTAL DETAILS**

We lithographically prepared 14 square, ferromagnetic particles of $Ni_{80}Fe_{20}$ (permalloy) of two different sizes 250 nm x 250 nm x 10 nm and 210 nm x 210 nm x 10 nm, 6 of the former and 8 of the latter. The manufacturing process follows Endean et al. [23, 24]. The samples were capped in Tantalum to reduce oxidation and no aging effects were noticed while samples were measured within a few weeks after manufacture. As in those works, nonmagnetic contacts were attached to the four corners of a square particle for constant DC, four terminal measurements. The direction of the magnetization was determined by measuring the anisotropic magnetoresistance (AMR) [25].

In zero applied magnetic field, the particles have some structure [23] but they are not large enough to support domain walls and are effectively single domain. They have four degenerate, magnetic ground states with the magnetization perpendicular to the sides of a square particle; these magnetic states are consistent with configurational anisotropy [23, 24, 26]. This results in the net magnetization either collinear or perpendicular to the current which results in a high or low resistance AMR state respectively. Application of a magnetic field along the diagonal of a particle removes the four-fold degeneracy giving a ground state with two degenerate minima with approximately perpendicular magnetizations; occupation of the specific minimum state can be determined by the different AMR resistances for the two orientations.

The energy barrier magnitude separating the two ground states is controlled by the magnitude of the applied field along the diagonal of the particle; increasing the applied field decreases the energy barrier. The zero-field barrier height, $U_o$, was measured using the same method as Endean et al.[23, 24] where the minimum field perpendicular to the magnetization direction needed to switch the magnetization direction, $H_{min}$, is

$$mH_{min} = \frac{U_o}{2} \qquad (2)$$

where $m$ is the magnetic moment of the particle using the magnetization value appropriate for permalloy. We should also point out that Endean et. al. [23] found the field dependence of the sample barrier heights was linear with magnetic field.

To obtain temperature records, our samples were mounted in a 4K cryostat that was cooled to 4K and allowed to warm to ambient temperature. The temperature was measured with a calibrated Cernox resistor attached to the sample holder. As the samples were warmed, an applied field along the diagonal of the sample was increased until two state switching was visually seen on an oscilloscope with dwell times on the order of 0.1 seconds to ensure that a sufficient number of switches were measured over our 10 second sample window. As the sample temperature further increased at this applied field, the dwell times became shorter until the switching between the two states was no longer discernable. At this point, a new data series was started by decreasing the field, thereby increasing the energy barrier, until two-state switching on the order of 0.1 seconds was achieved again. This allowed us to take multiple measurements on the same sample with different barrier heights.

Figure 1 is a semi-log plot which shows a typical measurement of dwell times for the two wells in a particle versus inverse temperature. Following Eq. 1, a fit was performed for both sets of data from each well to obtain the barrier height (slope of the line) and the prefactor (extrapolating to infinite temperature). The barrier heights measured by the slopes were consistent with those measured by the technique outlined by Endean et al[21]. In general, our data sets consisted of between 50 and 150 average dwell times each taken over times between 10 and 60 minutes; the data in Fig. 1 consists of about 60 measurements.

Given that the data are recorded as a time record, our data analysis is in terms of the dwell times in Eq. 1. The average dwell times are calculated from the time record for a given sample in a constant magnitude field applied in a specific direction as a function of temperature. In Fig. 1, we plot the results using the logarithm of Eq. 1, i.e.

$$\ln \tau = \ln \tau_o + \frac{U}{k_b T}. \quad (3)$$

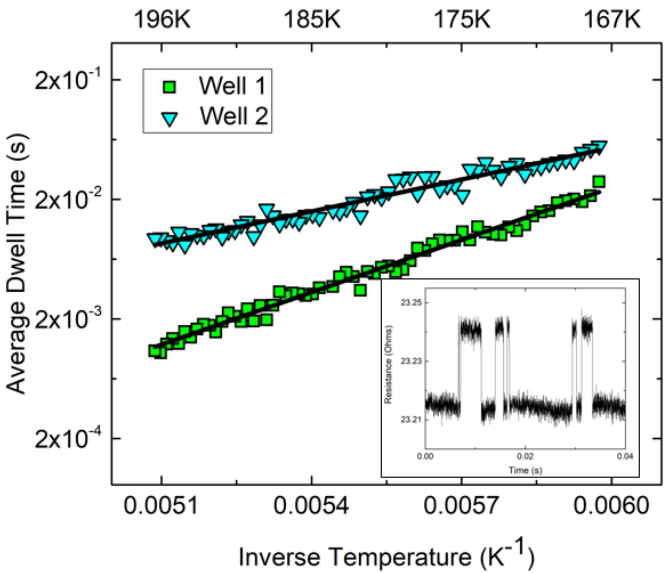

FIG 1. Average dwell times for the two wells in a particle undergoing two-state switching versus inverse temperature. The applied magnetic field is 132 Oe. Fits to Eq. 1 give both the prefactor, $\tau_o$, and the barrier height of the well. The insert is an example illustrating a short time segment of the resistance two-state switching measured as a function of time for the data collected at 196 K.

**RESULTS**

Figure 2 shows the measured prefactor as a function of the energy barrier over the 650 meV range in energy barriers. The surprise is prefactors, or $\tau_o$'s, as small as $10^{-28}$s were measured! These values are many orders of magnitude smaller than simply considering the prefactor to be a dwell time calculated by Brown's model [14] for ferromagnetic particles. One possible explanation is that the energy barriers are temperature dependent [27]. We removed this as a possible explanation by measuring the barrier at several temperatures for a number of samples using the method of Endean *et al* [20] which found some negligible temperature dependence. In addition, the data presented in Fig. 1 was analyzed using Eq. 1 with the form of $\tau_o$ being that calculated by Brown [14] that has both a $U$ and $T$ dependence (more about Brown's model later). By inverting his equation and solving for the barrier height as a function of temperature, we found that the barrier energy would need to change by 20% over ranges of about 10 K which is too large to explain the data. In addition, in many cases, the barrier height would have to increase for one well with increasing temperature while decreasing for the other which is, again, nonphysical. We want to stress the above clearly indicates a temperature dependence to the prefactor is not the cause for the prefactors we measure.

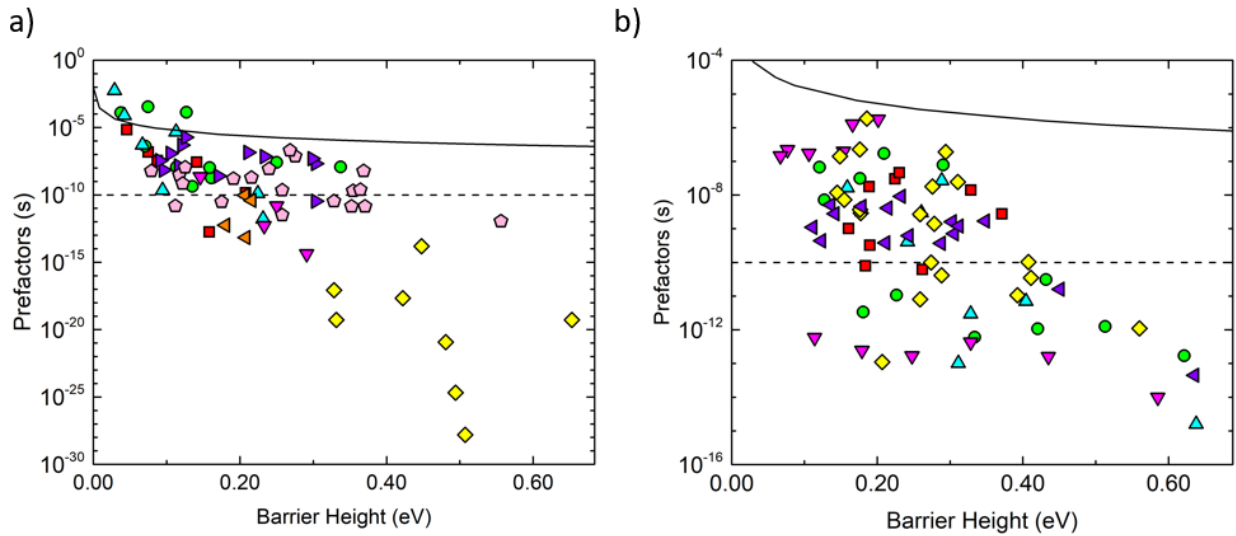

FIG 2. Prefactors versus Barrier Heights. Each symbol represents data for one of a) eight different 210 nm samples and b) six different 250 nm samples. By altering the applied magnetic field magnitude and/or the direction, each particle provides a number of data points. The black line is the expected prefactors for our particles from Brown's modeling discussed later and the dashed line is the time associated with FMR frequencies ($\sim 10^{-10}$ s). Replotting these data with different units on the abscissa as done later removes the large amount of scatter. Based on the fit of Eq 1, we find the errors in each prefactor value to be +/- 0.2 to 1.0 order of magnitude.

Both the large range in the prefactors and the unphysically short times, or more appropriately prefactors, indicate the general interpretation of $1/\tau_o$ or $f_o$ being an attempt frequency clearly is not correct. We find it is important to consider the ratio of the measured prefactors for a given particle given by

$$\frac{\tau_L}{\tau_H} = A_{PR} e^{\frac{U_L - U_H}{k_b T}}. \qquad (4)$$

where $U_L$ is the well with the smaller barrier height and $U_H$ is the larger barrier height energy when the two wells are uneven and that $A_{PR}$ is a phenomenological prefactor ratio added by the authors that should be unity if one considers simple detailed balance as first developed by Boltzmann [28] and expanded by Onsager [29, 30]. Here, we have assumed the characteristic dwell times, $\tau_o$, are the same for both wells. To compare our data in Fig. 2, we plotted the ratio of same-temperature data points from each well versus inverse temperature. As expected, the slope of the resulting line is equal to the difference in barrier heights over the Boltzmann factor but instead of a prefactor of unity as expected for simple detailed balance, we obtain a value near 800,000 for the data shown in Fig. 2. Values for $A_{PR}$ for all the particles at all the measured fields and temperatures were calculated from the data and except for the data with equal well energies, none had a detailed balance value of unity. This is a significant deviation from simple detailed balance.

In Fig. 3, we plot the values of $A_{PR}$ versus what we call the energy factor, EF; the rational for the form of the energy factor arises from the following logic.

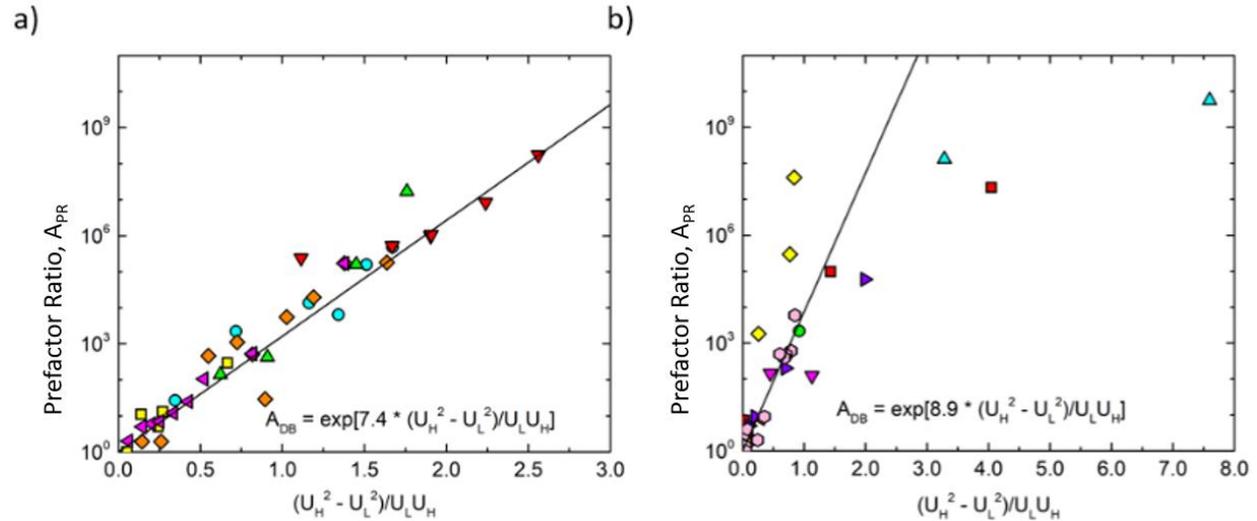

FIG. 3. The prefactor ratios, $A_{PR}$, versus the energy factor on a semilog plot of the two wells for the a) 250 nm samples[24] and the b) 210 nm samples. The lines are fits to an exponential with the constant, C, found to be 7.5 for the 250 nm particles and 8.9 for the 210 nm particles. For all detailed balance ratios, the well with the smaller barrier was divided by the well with the large barrier. These data and the fitting are discussed more thoroughly in the DATA ANALYSIS section.

The prefactor ratios, $A_{PR}$, can only depend on temperature and the barrier heights, the only physical properties of the particles that change. Since there was no discernable temperature dependence of $A_{PR}$, then the $A_{PR}$ must be a function only of the energy barriers. The form of the energy factor is a unitless term that goes to zero as the difference in energies decreases, required to give a prefactor of unity when the two wells are of equal depth as seen in Eq. 5.

$$EF = \frac{U_H^2 - U_L^2}{U_L U_H} \qquad (5)$$

**THEORETICAL ANALYSIS**

As we stated in the introduction, there have been a number of works to consider a prefactor that goes beyond a simple attempt frequency such as that by Brown [14]. To calculate the attempt frequency, Brown considered uniformly magnetized particles with uniaxial anisotropy. He used the Landau-Lifshitz-Gilbert (LLG) equation with an additional magnetic white noise term as his Langevin equation along with usual the assumption that the system barrier height was much larger than the thermal energies. Using his notation, the result for a symmetric, uniformly magnetized, uniaxial particle with a barrier height larger than the thermal energies is

$$f \approx \frac{2U^{\frac{3}{2}}}{\sqrt{\pi}\,\tau_N (k_b T)^{\frac{3}{2}}} e^{-U/k_b T}, \quad (8)$$

where $U$ is the energy barrier, $T$ is the temperature, and $\tau_N$ is known as the Neel time [14] and has the form

$$\tau_N = \frac{V M_s (1+\alpha^2)}{2\gamma \alpha k_b T}, \quad (9)$$

where $V$ is the volume of the particle, $M_s$ the saturation magnetization of the particle, $\alpha$ is the unitless Gilbert damping factor which replaces $\beta$ for magnetic systems, and $\gamma$ is the gyromagnetic ratio. When discussing magnetic systems, $\alpha$ is more appropriate to use instead of the damping rate, $\beta$, since the unitless Gilbert damping factor is measureable by ferromagnetic resonance (FMR) [31, 32]. For nanoscale sized particles, the values of $f_o$ can range from $10^9 - 10^{11}$ Hz which is close to FMR frequencies and usually researchers assume $f_o$ to be the FMR frequency [14]. The value obtained using Eq. 8 is the basis of the solid lines in figure 2 that clearly do not describe the data.

As mentioned in the Introduction, others have considered more complex situations than single domain particles with uniaxial anisotropy [3, 7, 8, 9, 10, 12, 13, 15, 16, 17, 18 20, 21, 33]. In general, these extensions can be thought of as considering entropic additions to the prefactor such as is often done in chemical systems by deriving a transition rate using the Gibbs Free Energy which results in the Eyring equation[34]. The review paper by Coffey and Kalmykov [19] consolidated the work of others when it came to applying multidimensional solutions to physical systems which have multiple degrees of freedom [8, 9, 10, 19, 20, 21, 33]. Of these, the one we found useful to describe our results is Talkner's model using a "mean first passage time" approach to many particles in a well. He derived a solution that can roughly be described as the ratio of how many particles are near the saddle point and near the bottom of the well [10]. In other words, the more particles near the saddle point relative to the minimum, the faster the system will transition. He took the number of particles near the saddle point and well minimum as proportional to the product of eigenvalues of the Hessian matrix [35] when evaluated at those two states respectively.

Using a "mean first passage time" approach [10, 19, 22], Talkner obtained the result

$$f \propto \frac{1}{2\pi} \frac{\prod_i^N \lambda_i^m}{\prod_i^N |\lambda_i^s|} e^{-\Delta U/k_b T}, \quad (10)$$

where $\lambda^m$ and $\lambda^s$ are the eigenvalues of the well minimum and saddle point respectively and $i$ designates each normal mode. To obtain Eq. 10, Talkner calculated the probability of the system being near the well minimum, $P_m$, and near the saddle point, $P_s$, as

$$P_m \propto \left(\prod_i^N \lambda_i^m\right)^{-1}; \; P_s \propto \left(\prod_i^N \lambda_i^s\right)^{-1}. \quad (11)$$

Equations 10 and 11, taken in combination, can be thought of physically as the higher the probability the system is near the well minimum, the more time it needs to transition over the barrier and the transition frequency decreases. Similarly, the higher the probability the system is near the saddle point, the less time the system needs to transition and the transition frequency increases. In a very low damping regime, which Talkner assumed, the particles in the system keep much of their excess energy from each transition. This results in the particles spending an appreciable amount of their time near the saddle point instead of near the well minimum. Because the particles are near the saddle point, they have a higher chance of transitioning back over the well early.

We now consider the prefactor ratio or detailed balance that was first described by Boltzmann in 1872 as an extension of microscopic reversibility [36]. In his picture, a system has a probability of transitioning from a state $i$ into a state $j$, $P_{i \to j}$, separated by an energy barrier, $U_{i \to j}$, with the probability given by

$$P_{i \to j} = e^{U_{i \to j}/k_b T}, \quad (12)$$

and similarly, a probability of transitioning from a state $j$ into a state $i$, $P_{j \to i}$, such that

$$\frac{P_{i \to j}}{P_{j \to i}} = e^{(U_{i \to j} - U_{j \to i})/k_b T}, \quad (13)$$

assuming $f_o$ is the same for both wells. This assumption is based on the prefactor from Eq 4 depending on $U^{3/2}$ which will only vary by an order of magnitude at most over the range in barrier heights measured.

For a simple physical picture of why the prefactor is different from a simple dwell time or attempt frequency, one can use the Helmholtz free energy in the Boltzmann factor,

$$F_i = U_{i \to j} - TS_i, \quad (14)$$

where $S_i$ is the entropy of the state, $i$ [36]; this alteration is similar to the use of the Gibbs free energy in the Arrhenius equation by chemists [34]. By replacing the barrier height in Eq. 12 with the free energy from Eq. 14, we obtain an additional prefactor term, $\exp(-S_i/k_b)$. This extra entropy term can be thought of as a measure of the many possible paths over the barrier; the number of paths increases with the increase in the barrier. Although not explicitly stated as the use of the Helmholtz free energy, Yelon and Movaghar [38] explain that these extra paths allow a combination of low energy, thermal phonons to act collectively to cause a state switch instead of a single phonon of energy on the order of the barrier height. Since the entropy does not have to be the same for both states, an additional term in the prefactor ratio can survive changing Eq. 13 to

$$\frac{P_{i \to j}}{P_{j \to i}} = e^{S_j - S_i} e^{(U_{i \to j} - U_{j \to i})/k_b T}. \quad (15)$$

In the case where the configuration entropy of the two states is the same, the detailed balance ratio is again unity.

By comparing Eq. 15 to Eq. 10, we obtain a configurational entropy term[37] that goes as

$$S = \ln \frac{\prod_i |\lambda_i^s|}{\prod_i \lambda_i^m} + const. \quad (16)$$

which is similar in form to Boltzmann's entropy formula [39] where the number of microstates is now given by the ratio of the products of the eigenvalues.

ANALYSIS OF EXPERIMENTAL RESULTS

Starting with a determination of the Hessian matrices needed for our analysis in Eqs 9 and 10, we start by considering the magnetostatic energy of the system given by

$$E(\theta_1, \ldots, \theta_N) = E_{exchange}(\theta_{n.n.}) + E_{dipole-dipole}(\theta_1, \ldots, \theta_N) + E_{Zeeman}(\theta_1, \ldots, \theta_N), \quad (17)$$

where each term is dependent upon all the degrees of freedom (orientations of each macrospin). Note, since our system is permalloy, we omitted the crystalline anisotropy energy. The Hessian matrix, $K$, for a system is the matrix of all the second derivatives with respect to a system's degrees of freedom included in Eq. 17 is given by

$$K_{ij} = \frac{\partial^2 E}{\partial \theta_i \partial \theta_j}. \quad (18)$$

where $\theta_{i,j}$ are the polar angles of the macrospins of the system. This results in a matrix that is of size $N^2 \times N^2$ where N is the number of macrospins in the system. It provides the curvature of the energy landscape for any generalized magnetization state (unique configuration of all the macrospins in the system). Once the Hessian matrix is determined in terms of the generic angles, $\theta$, the macrospin angles can be found using simulations of the magnetization for the case when the net magnetization is pointing towards a well minimum and when pointing towards the maximum, that for multidimensional systems tends to be a saddle point. In our implementation, the macrospins are approximated as the unit cells of the magnetization simulations performed using the LLG Micromagnetics Simulator [40] where the angles of the magnetization of each unit cell are used as the angle of each macrospin's magnetization.

For these simulations, we considered a particle of size $250\ nm \times 250\ nm \times 10\ nm$ divided into $50 \times 50 \times 2$ unit cells with a $M_s$ of $800\ emu/cm^3$, and an exchange stiffness constant of $1.05\ \mu erg/cm$ as appropriate for permalloy [41]. This cell size is approximately that of the exchange length in permalloy. An external field of varying magnitude was directed along the diagonal of the particle. The zero temperature simulations were initiated by saturating the particle along the diagonal and then returning the field to zero. When the net magnetization was pointing along the field direction, we referred to this as the barrier direction. When the net magnetization relaxed to a different direction, we define this as the minimum well direction.

The angle of each macrospin at either the minimum state or the barrier state $\{\theta_1, \theta_2 \ldots \theta_N\}$ is placed into the Hessian matrix which is then diagonalized giving both the eigenvectors, the normal modes of the system, and the eigenvalues, $\lambda$. Since a particle has multiple degrees of freedom, there will exist many normal modes that are independently attempting to transition over the barrier which can be related to the configurational entropy which we will discuss.

Returning now to the data in Figure 3, we need to develop a form for the detailed balance prefactor, $A_{PR}$. First, the only relevant physical properties of a given set of data for a particle are the well energies and their temperature but since $A_{PR}$ was not temperature dependent, only the well energies were considered relevant. Second, if the barrier heights are the same size for each well, $A_{PR}$ must be unity, returning to the expected detailed balance. And lastly, this prefactor is expected to be symmetric when exchanging the subscripts $L$ and $H$. Since $A_{DB}$ goes to unity as the difference in well heights goes to zero, we expect $A_{DB}$ to have some form of exponential dependence on the energies. Combining these we find a suitable $A_{DB}$ given by

$$A_{PR} \approx e^{C(U_H^2 - U_L^2)/U_L U_H}, \quad (19)$$

where $C$ is a fitting parameter. The values for $A_{DB}$ for all the samples are plotted in Fig. 3 versus what we defined as the energy factor, $(U_H^2 - U_L^2)/U_L U_H$. The energy factors for all $A_{PR}$ are positive by our convention of dividing the higher energy state (small barrier) dwell times by the lower energy state (large barrier) dwell times. With this added term, the Arrhenius law for a double well system with wells labelled $L$ and $H$, becomes

$$\tau_L = \tau_o e^{C \frac{U_H}{U_L}} e^{\frac{U_L}{k_b T}}, \quad (20)$$

where $U_L$ is the height of the barrier for the smaller well and $U_H$ is the height of the barrier for the larger well for consistency. If Eq. 20 for both $\tau_L$ and $\tau_H$ are taken in ratio, the ratio will be Eq. 4 with the form of $A_{PR}$ in Eq. 19.

As can be seen in Fig. 3b, there is one data point that deviates from this simple exponential form for large energy differences between the wells for the 210 nm particles. It is not clear if this is a bad data point or indicates real behavior but it can be replicated as discussed later and is given for completeness. Note that, Eq. 16 can only have values greater than one for our values of $C$, meaning that Eq. 16 can only increase dwell times and therefore cannot explain the measured prefactors that are smaller than the expected values from Brown. However, the form of $A_{PR}$ is more or less consistent with the data shown in Fig. 5. A more appropriate form of $A_{PR}$ that explains the behavior seen in Fig. 3 will be presented later but first we discuss results from our simulations.

The deviation from linearity has been replicated by our simulations of thermally activated, two-state switching of particles we calculated using Mumax3 [42]. For these simulations, we used 85x85x2 unit cells and the same material parameters as used for the LLG Micromagnetics simulations [32, 33, 40]. Time records of the resistance were obtained for 4 $\mu s$ at temperatures of 300 K, 290 K, and 280 K with applied fields of around 60 Oe. The magnetic field direction was slightly off diagonal by as much as 0.3º to create uneven wells. From the simulated time records exhibiting RTN, $A_{PR}$ was calculated for over twenty-five different field strengths and directions; the $A_{PR}$ values from the simulations are shown in Fig. 4. As seen, the simulations replicate the general form of the data. The fitting constant, $C$, is smaller than what is seen experimentally which goes hand in hand with the smaller range in values of $A_{PR}$; these are due to the simulations having smaller barrier heights which was required to match the smaller time records.

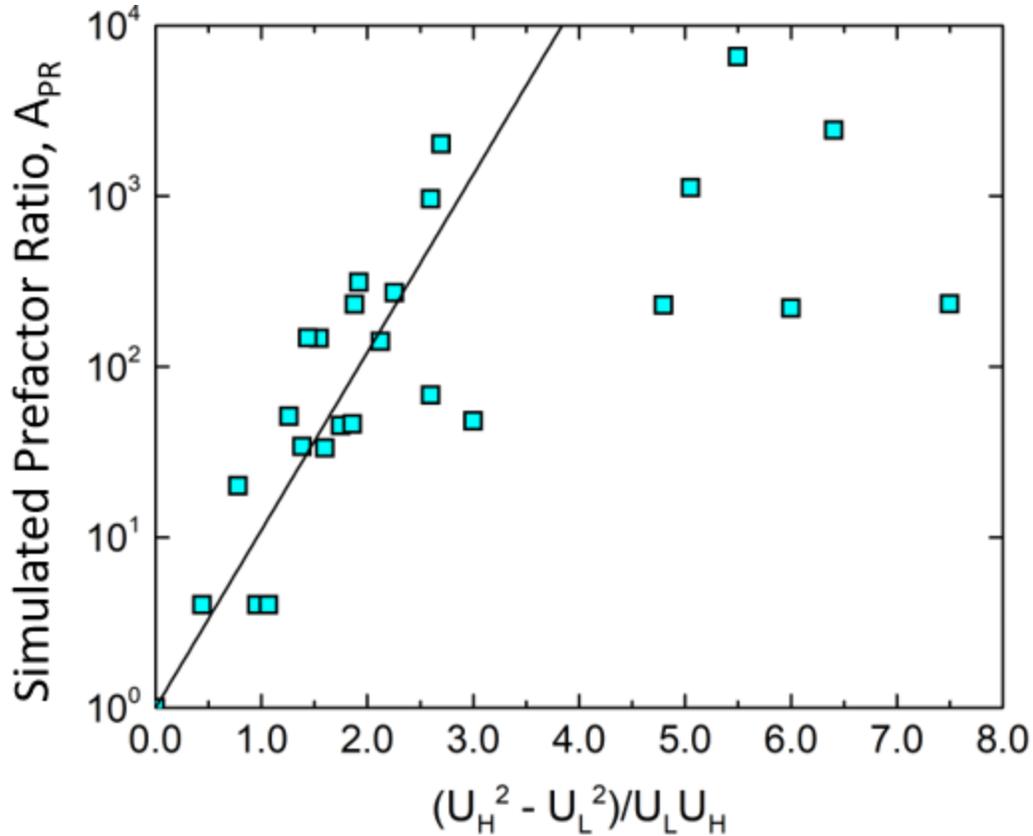

FIG. 4. Simulation determined detailed balance prefactors, $A_{DB}$, versus the energy factor. The fitted line is the exponential given in the plot performed up to an energy factor of 3.0 before the points start to curve over. The fitting constant, $C$, from Eq. 19 for the curve fit of the data up to the energy factor of 4.0 is 2.4.

The prefactor ratio, $A_{PR}$, decreases as the energy difference between the wells gets smaller and $A_{PR}$ decreases becoming unity when the barrier heights are the same for both wells. Looking at the entropy prefactor from Eq. 15, as the difference in entropy between the two states becomes larger, the prefactor also becomes larger. This suggests that the configurational entropy of a state increases with the height of the barrier for that state.

Due to our particles being at the mesoscale, we must account for the extra degrees of freedom the system has by using a multidimensional solution. This will provide for a more correct solution to both the dependence of the prefactors on the barrier height and the detailed balance problem both seen in Fig. 3. Based on the $50 \times 50$ unit cells from our LLG Micromagnetics simulations, we obtained 2,500 eigenvalues for each state that varied in value from -0.3 eV to 5.5 eV. Using these eigenvalues and Eq. 10, we calculated the prefactors for Talkner's result with one alteration. We limited the eigenvalues for both the minimum state and saddle point state used in Eq. 10 to those eigenvalues that are less than or equal to the height of the barrier. We point out that if Talkner's proposal that the maximum energy is not limited to the barrier energy, it would result in values of $A_{DB}$ that decrease with an increasing energy factor and this is inconsistent with our data (see Appendix A for details).

In Fig. 5, we show the measured prefactor ratios along with the results of our slightly altered Talkner's model where the symbols connected by lines designate keeping three different values of $\left(\prod_i \lambda_i^{minimum}\right)_H$ constant and allowing the results for the smaller well to vary i.e. keeping the larger well constant and decreasing the size of the smaller well as you move to the right. The plateaus are an artifact of the granularity of the cells used in our simulations that was limited by the time required for the calculations.

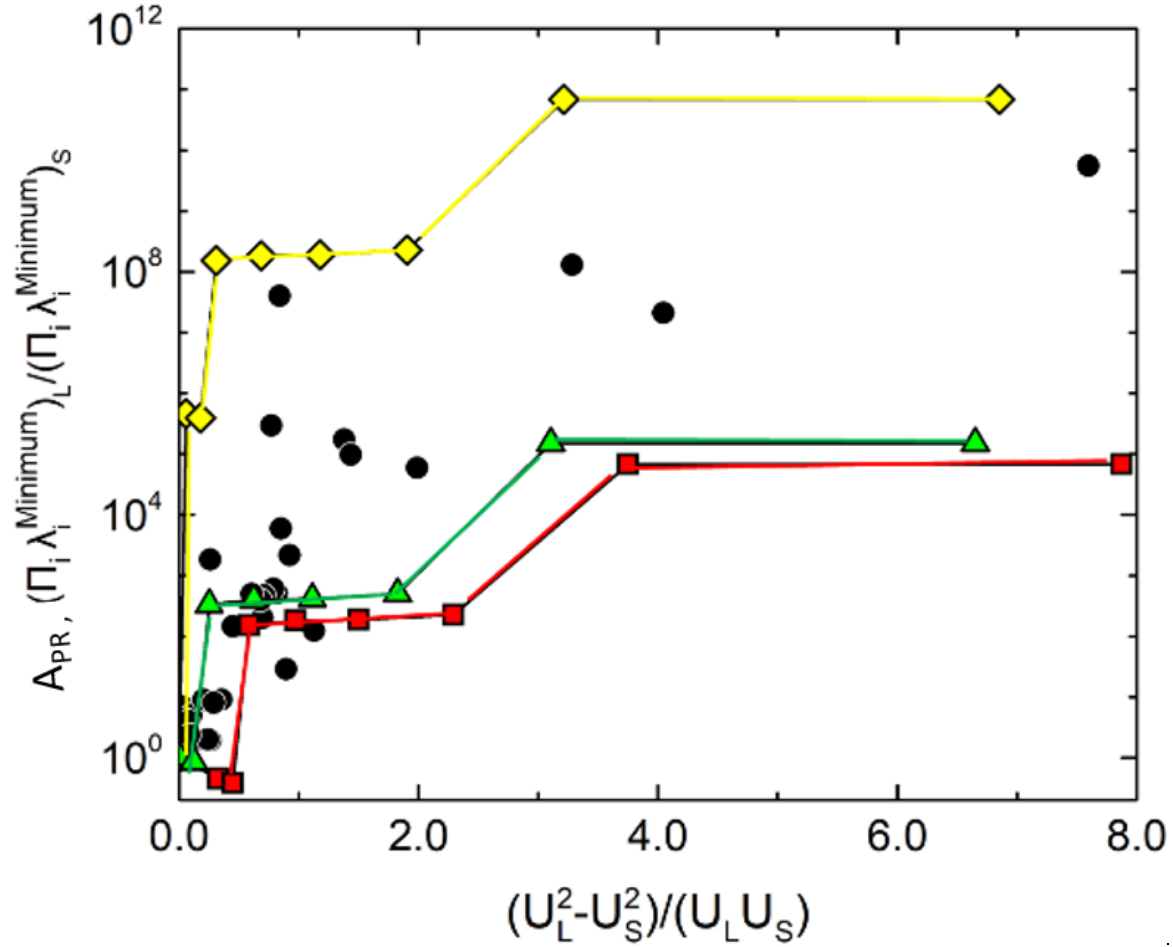

FIG. 5. The detailed balance prefactors plotted versus the energy factor for both our measured data, $A_{PR}$ (black dots) and for our theoretical prefactor results (shapes with lines) obtained using Eq. 8. The three different lines designate keeping three different $\left(\prod_i \lambda_i^{minimum}\right)_H$ constant and allowing the results for the smaller well to vary.

Our model's results show a similar trend to our measurements along with a similar change in orders of magnitude. We also performed the same analysis to LLG simulation data of two state switching which showed similar results as our measured data.

SUMMARY

In conclusion, in the almost 80 years since Kramers' work, a rigorous experimental investigation of the Arrhenius prefactor, $\tau_o$, has been difficult since these measurements themselves are difficult. To date, Arrhenius fits of dwell times have been made with only a few data points[15, 16, 43, 44]. Also, it is common in these other works to rely on intrinsic manufacturing defects such as sample size or shape irregularities to change the height of the energy barrier to probe $\tau_o$. In our model system however, we have been able to collect a significant amount of data while precisely controlling the barrier height using an applied field. These two facts have allowed us to study the physics behind $\tau_o$ more extensively as shown in Figs. 2 and 3.

With this precise control and extensive collection of data, we have shown that we obtain Arrhenius law prefactors that are orders of magnitude different than the simple interpretation of them being an attempt frequency would predict. Using Talkner's model, limited to eigenvalues of energies smaller than the height of the barrier, we can explain the observed divergence from simple detailed balance as we show in Fig. 5. In this we show an additional prefactor term that varied by 10 orders of magnitude over an energy factor range of eight. This has far reaching consequences beyond physics research as other fields of study such as chemistry, biology, and even qubit physics requires an understanding of the dynamics of two-state switching in mesoscale systems.

A very simple way to consider how the Arrhenius equation is altered for mesoscale systems is to replace the internal energy, $U$, in the Boltzmann factor with the Helmholtz free energy, $U - TS$ [37]. The entropy term, divided by $k_b T$, becomes a temperature independent exponential prefactor multiplied by the dwell time or attempt frequency. We believe that this additional prefactor due to the entropy is the same as our phenomenological detailed balance prefactor which gives us the form

$$A_{PR} = e^{C(U_H^2 - U_L^2)/U_L U_H} = \frac{\prod_i |\lambda_i^s|}{\prod_i \lambda_i^m} = e^{S_j - S_i}. \quad (21)$$

Thus our results are a direct measurement of the entropic contribution to the Arrhenius Law prefactor.

Although there are questions remaining, it is important to note we have provided evidence that the often-used Arrhenius equation prefactor for mesoscale systems, including physical, chemical, and biological systems, can be order of magnitude different than expected for a simple attempt frequency. In addition to an understanding of the physics of the prefactor, the prefactor can be sufficiently different to produce errors in the Boltzmann factor used to determine the energy barriers. In simple terms, the prefactor of mesoscale systems must include entropic considerations. This possibility has been long considered as evidence by the numerous theoretical models but testing of the models has been absent until our development of an ideal model system with all the relevant energies measured independent of the relevant data for the prefactor determination.

We would like to thank Kirill Belashchendo, Mark Stiles, Rafael Fernandes, Michael Flatte, Alex Kamenev, Allen MacDonald, Ray Orbach, Ralph Skomski, Mark Stiles, Evgeny Tsymbal, and Randy Victora for their valuable conversations which helped with this work. This work was supported primarily by NSF Grant No. DMR 1609782 and DMR 2103704. Portions of this work were conducted in the Minnesota Nano Center, which is supported by the National Science Foundation through the National Nano Coordinated Infrastructure Network (NNCI) under Award Number ECCS-1542202. The data that support the findings of this study are openly available at the Data Repository for the University of Minnesota (DRUM) at https://doi.org/10.13020/9d4f-d239.

**References**

[1] S. Arrhenius, "Über die Reaktionsgeschwindigkeit bei der Inversion von Rohrzucker durch Säuren." *Z. Phys. Chem.* **4** 226-248 (1889).

[2] S.R. Logan, "The origin and status of the Arrhenius equation." *J. Chem. Educ.* **59**, 279 (1982).

[3] H.A. Kramers, "Brownian motion in a field of force and the diffusion model of chemical reactions." *Physica* **7,** 284 (1940).

[4] C.M. Dobson, A. Šali, and M. Karplus. "Protein folding: a perspective from theory and experiment." *Angew. Chem. Int. Ed. Engl.* **37**, 868 (1998).

[5] G.R. Fleming, S.H. Courtney, and M.W. Balk. "Activated barrier crossing: Comparison of experiment and theory." *J. Stat. Phys.* **42**, 83 (1986).

[6] R.A. Marcus, "Electron transfer reactions in chemistry. Theory and experiment." *Pure Appl. Chem.* **69** 13 (1997).

[7] Mel'nikov, V. I., and Meshkov. S.V. "Theory of activated rate processes: Exact solution of the Kramers problem." *The Journal of chemical physics* **85**, 1018 (1986).

[8] G.H. Vineyard, "Frequency factors and isotope effects in solid state rate processes." *J. Phys. Chem. Solids* **3** 121 (1957).

[9] M. Borkovec, J.E. Straub, and B.J. Berne, "The influence of intramolecular vibrational relaxation on the pressure dependence of unimolecular rate constants." *J. Chem. Phys.* **85**, 146 (1986).

[10] P. Talkner, "Mean first passage time and the lifetime of a metastable state." *Z. Phys. B.* **68**, 201 (1987).

[11] P. Hänggi, F. Marchesoni, and P. Grigolini. "Bistable flow driven by coloured Gaussian noise: a critical study." *Z. Phys. B* **56**, 333 (1984).

[12] A.J. Bray, and M.A. Moore. "Influence of dissipation on quantum coherence." *Phys. Rev. Lett.* **49**, 1545 (1982).

[13] M. Ghasemi, N. Balar, Z. Peng, H. Hu, Y. Qin, T. Kim, J.J. Rech, M. Bidwell, W. Mask, I. McCulloch, W. You, A. Amassian, C. Risko, B.T. O'Connor and H. Ade, "A molecular interaction–diffusion framework for predicting organic solar cell stability. *Nat. Mater.* **20,** 525

[14] (2021).W.F. Brown Jr, "Thermal fluctuations of a single-domain particle." *Phys. Rev.* **130.5**, 1677 (1963).

[15] S. Krause, G. Herzog, T. Stapelfeldt, L. Berbil-Bautista, M. Bode, E.Y. Vedmedenko, and R. Wiesendanger, "Magnetization reversal of nanoscale islands: How size and shape affect the Arrhenius prefactor." *Phys. Rev. Lett.* **103.12**, 127202 (2009).

[16] M. Respaud, M. Goiran, J.M. Broto, F. Lionti, L. Thomas, B. Barbara, T.O. Ely, C. Amiens, and B. Chaudret, "Dynamical properties of non-interacting Co nanoparticles." *Europhys. Lett.* **47.1**, 122 (1999).

[17] A.N. Cleland, J.M. Martinis, and J. Clarke. "Measurement of the effect of moderate dissipation on macroscopic quantum tunneling." *Phys. Rev. B* **37**, 5950 (1988).

[18] J. Schroeder, and J. Troe, "Elementary reactions in the gas-liquid transition range." *Annu. Rev. Phys. Chem.* **38**, 163 (1987).

[19] W.T. Coffey, Y.P. Kalmykov, "Thermal Fluctuations of Magnetic Nanoparticles: Fifty Years After Brown." J. Appl. Phys. **112**, 121301 (2012).

[20] M. Borkovec, Ph.D. thesis, Columbia University, New York (1986).

[21] E. Pollak, H. Grabert, and P. Hänggi. "Theory of activated rate processes for arbitrary frequency dependent friction: Solution of the turnover problem." *J. Chem. Phys.* **91**, 4073 (1989).

[22] P. Hänggi, P. Talkner, and M. Borkovec. "Reaction-rate theory: fifty years after Kramers." *Rev. Mod. Phys.* **62**, 251 (1990).

[23] D.E. Endean, C.T. Weigelt, R.H. Victora, E. Dan Dahlberg, "Measurements of configurational anisotropy in isolated sub-micron square permalloy dots." *Appl. Phys. Lett.* **103**, 042409 (2013).

[24] D.E. Endean, C.T. Weigelt, R.H. Victora, and E. Dan Dahlberg. "Tunable random telegraph noise in individual square permalloy dots." *Appl. Phys. Lett.* **104**, 252408 (2014).

[25] T. McGuire, and R.L. Potter. "Anisotropic magnetoresistance in ferromagnetic 3d alloys." *IEEE Trans. Magn.* **11**, 1018 (1975).

[26] R.P. Cowburn, A.O. Adeyeye, and M.E. Welland. "Configurational anisotropy in nanomagnets." *Phys. Rev. Lett.* **81**, 5414 (1998).

[27] M. Stier, A. Neumann, A. Philippi-Kobs, H.P. Oepen, and M. Thorwart. "Implications of a temperature-dependent magnetic anisotropy for superparamagnetic switching." *J. Magn. Magn. Mater.* **447**, 96 (2018).

[28] L. Boltzmann, "Weitere Studien über das Wärmegleichgewicht unter Gasmolekülen." *Sitzungsberichte Akademie der Wissenschaften*, **66**, 275-370 (1872).

[29] L. Onsager, "Reciprocal relations in irreversible processes I." *Phys. Rev.* **37**, 405-426 (1931).

[30] L. Onsager, "Reciprocal relations in irreversible processes II." *Phys. Rev.* **38**, 2265-2279 (1931).

[31] Y. Tserkovnyak, A. Brataas, and G.E.W. Bauer, "Enhanced Gilbert damping in thin ferromagnetic films." Phys. Rev. Lett. **88**, 117601 (2002).

[32] M. Oogane, M., T. Wakitani, S. Yakata, R. Yilgin, Y. Ando, A. Sakuma, and T. Miyazaki, "Magnetic damping in ferromagnetic thin films." *Jpn. J. Appl. Phys.* **45**, 3889 (2006).

[33] H.C. Brinkman, "Brownian motion in a field of force and the diffusion theory of chemical reactions. II." *Physica* **22**,149 (1956).

[34] H. Eyring, "The activated complex in chemical reactions." *J. Chem. Phys.* **3.2**, 107-115 (1935).

[35] L.E. Spence, A.J. Insel, and S.H. Friedberg. *Elementary linear algebra*. Prentice Hall, 2000.

[36] L. Boltzmann, *Lectures on gas theory*. (Courier Corporation, 2012).

[37] L. Desplat and J. Kim. "Entropy-reduced Retention Times in Magnetic Memory Elements: A Case of the Meyer-Neldel Compensation Rule." *Phys. Rev. Lett.* **125**, 107201 (2020).

[38] A. Yelon and B. Movaghar, "Microscopic Explanation of the Compensation (Meyer-Neldel) Rule, *Phys. Rev. Lett.* **65**, 619 (1990).

[39] L. Boltzmann, *Vorlesungen über Gastheorie,* (J.A. Barth, Leipzig 1896), Vol. I.

[40] M.R. Scheinfein and E.A. Price. "LLG User Manual v2. 50." *Code of the LLG simulator can be found at http://llgmicro.home.Mindspring.com* (1997).

[41] R. O'Handley, "Modern Magnetic Materials" (Wiley Interscience, 2000).

[42] A. Vansteenkiste, J. Leliaert, M. Dvornik, M. Helsen, F. Garcia-Sanchez, B. Van Waeyenberge, "The design and verification of MuMax3." *AIP Adv.* **4**, 107133 (2014).

[43] S.R. De Groot and P. Mazur. *Non-equilibrium thermodynamics*. Courier Corporation, 2013.

[44] F. Liu, Y. Ding, R. Kemshetti, K. Davies, P. Rana, and S. Mao, "Electrical low frequency random telegraph noise in magnetic tunnel junctions." *J. Appl. Phys.* **105**, 07C927 (2009).

**Appendix A**

Using our eigenvalues obtained from our simulated results and Eq. 10, we calculated the prefactors for Talkner's result as a function of field and obtained trends for both the ratios of the products of eigenvalues at the saddle point and bottom of the well (Fig. 6a) which should be equivalent to the prefactors and the ratio of the ratio of the prefactors which are equivalent to the prefactor ratio, $A_{PR}$ (Fig. 6b). It is important to note that the prefactors in Fig. 6a increase with decreasing field strength which is equivalent to increasing barrier height. This result is the inverse of what we see in the measured data of Fig. 2.

We also calculated the ratio of the prefactors for both wells at a given field and temperature which should give similar results to the detailed balance prefactor, $A_{PR}$, in Fig. 3. For every combination of prefactors from Fig. 6a, we divided the prefactor for the more shallow energy well, $\tau_{LS}$, by the prefactor for the deeper energy well, $\tau_L$, to ensure that this ratio is consistent with the measured ratios and the results shown in Fig. 6b. Each line represents holding the deeper energy well's prefactor constant and changing the more shallow well's prefactor. The ratio of these prefactors decrease by many orders of magnitude with increasing energy factor which, again, is the opposite trend of the measured data as seen in Fig. 3.

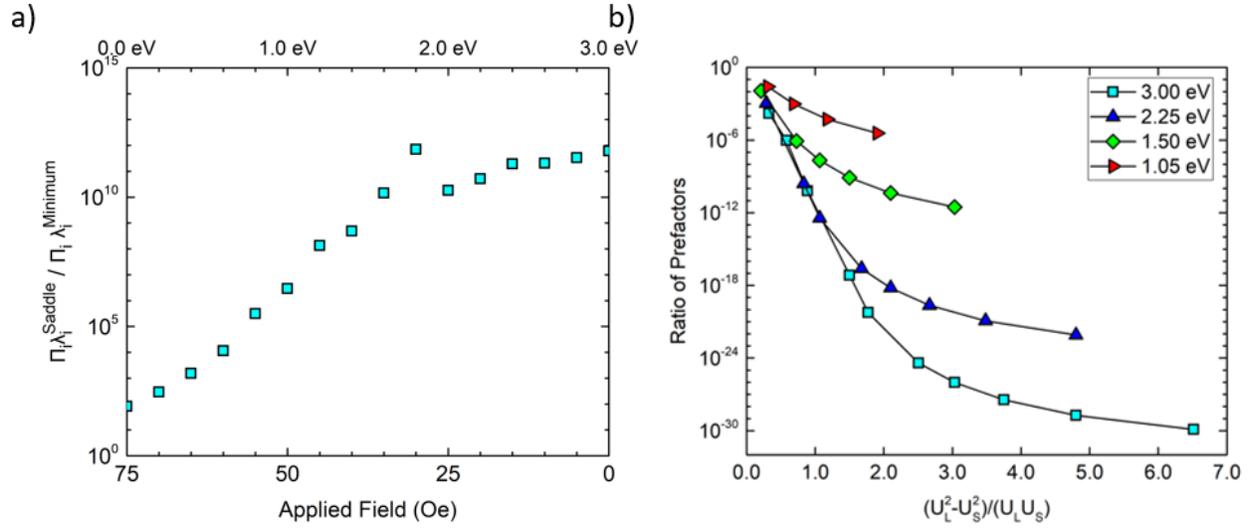

**Figure 6.** A plot of a) the ratio of products of eigenvalues from Eq. 9 versus applied field on a semilog plot which should be equivalent to the prefactors versus barrier height and b) the ratio of the prefactors from a) using simulation data. The lines in b) represent keeping $\tau_H$ constant and changing $\tau_L$.